# Suppression of the emittance growth induced by CSR in a DBA cell

CUI Xiao-Hao(崔小昊)[1)]　　JIAO Yi(焦毅)　　XU Gang(徐刚)　　HUANG Xi-Yang(黄玺洋)

Institute of High Energy Physics, Chinese Academy of Sciences, Beijing 100049, China

**Abstract:** The Emittace growth induced by Coherent Synchrotron Radiation(CSR) is an important issue when electron bunches with short bunch length and high peak current are transported in a bending magnet**.** In this paper, a single kick method is introduced which could give the same result as the R-matrix method, and much easier to use. Then with this method, an optics design technique which could minimize the emittance dilution within a single achromatic cell.

**Key words**: coherent synchrotron radiation, emittance, lattice design, DBA

## 1. Introduction

The design and study on the next generation light source based on Energy Recovery Linac (ERL) and Free Electron Lasers (FEL) have been proposed worldwide. In these machines, electron bunches with short bunch length, high charge and small emittance are generated and transported and it is very important to minimize the transverse emittance growth in order to achieve high quality electron beams. Emission of Coherent Synchrotron Radiation (CSR) is considered to be one of the most critical sources of the beam emittance dilution when bending magnets or bunch compression sections exist in the transport lines.

According to the CSR wake potential[1], the energy change of a electron due to the CSR emission is a function of its longitudinal position in the bunch. As a consequence, different bunch slices are deflected with different angles in a bending magnet, and this deflecting error dilutes the projected emittance of the electron bunch[2]**.** This effect has been studied intensively[3,4], and it is shown that if the longitudinal electron distribution doesn't change significantly, the rms energy spread caused by CSR can be estimated by the function:

$$\Delta E_{rms} = 0.22 \frac{eQL_b}{4\pi\varepsilon_0 \rho^{2/3} \sigma_s^{4/3}} \quad , \tag{1}$$

Where e is electron charge, Q is the bunch charge, $L_b$ is the length of the bending magnet, $\rho$ is the bending radius and $\sigma_s$ denotes the rms bunch length. For constant bending radius, it is a linear function of s, the longitudinal path length, and can be simplified as $\triangle E = \kappa s$, where k denotes the coefficient of $L_b$ in function (1) . Under this linear approximation, two optics design techniques have been introduced for the suppression of the CSR induced emittance growth: The envelope matching[5], and the cell-to-cell phase matching[6]. However, all these two methods have their shortcomings, envelope match can only minimize instead of completely cancelling the emittance growth, and cell-to-cell phase matching method relies strongly on the symmetrical character of the lattice, and cell-to-cell betatron phase advance. In the present paper, another method of emittance cancellation is shown, which can completely cancel the emittance growth due to linear CSR effect within a single achromatic cell.

The paper is organized as follows. In section 2, a single kick method is introduced to describe the transverse effect of CSR. In his paper [7], R. Hajima proposed a 5-by-5 R-matrix to calculate the emttance growth arising from CSR effect. Using this matrix method, an exact description of the linear CSR effect is acquired and it can be treated as standard transfer matrix. But all calculations involve complex 5-by-5 R-matrix manipulations and it's hard to get an intuitive idea of what happened. For this reason, S. Di. Mitri[8] used a single kick to describe the CSR induced transverse effect. But his kick method can only give a qualitative rather than quantitative agreement with the R-matrix method. We will give a single kick description which gives the same results as the R-matrix method, and avoids complicated matrix manipulations. In section 3, the CSR induced emittance in a DBA cell is studied using this single kick method, and a constraint condition on the lattice is got which could cancel the CSR kick completely. In section 4, different DBA lattices with various beam and optical parameters are simulated using code ELEGANT[9], and it is verified that the DBA lattice meets our condition really gives the smallest emittance growth.

## 2. Kick description of CSR effect

### 2.1 Kick description of the dispersion

The matrix description of dispersion has been widely known and easy to use. Consider a particle with energy deviation $\delta$, whose transverse coordinates are $(x_0, x_0')$ at the entrance of a dipole. When it goes through the dipole, its coordinates change not only by the transfer matrix of the dipole, but also, a dispersion term is added. If we are not interested in the particle information inside the magnet, the dispersion can be considered as a point kick at the middle point of a dipole magnet. To get the same result with the matrix description, we choose the middle point kick to get the same result at the exit of a dipole, that is $M_{half}(x_{kick}, x_{kick}')=(D, D')\delta$, where $M_{half}$ is the 2-by-2 transfer matrix of half a dipole, and $(D, D')$ is the dispersion function generated by the dipole. From this equation, dispersion kicks can be calculated. Take a sector dipole as an example, the dispersion function at the end of the magnet is $(D, D')=(\rho(1-\cos\theta), \sin\theta)$. From the above equation the dispersion kick can be calculated as $(x_{kick}, x_{kick}')=(0, 2\sin(\theta/2))\delta$, where $\theta$ is the bending angle of the dipole. We can see that by placing the kick at the middle point of the dipole, the expression of a kick is simplified. Later, when we study beam dynamics in a normalized coordinates, this advantage will become more obvious.

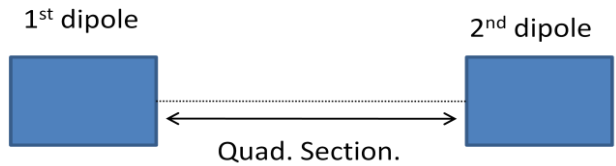

Fig. 1. Layout of a double-bending cell

To show the usefulness of this kick description, we'll talk about the achromatic condition of a double bending cell. The double bending cell is composed of two bending dipoles at the head and end of the cell separately and drifts and quadrupoles between, as shown in Fig. 1. Those two dipoles have the same bending directions but may have different bending radius and bending angles. In the discussion below, we'll use the normalized phase space coordinates $(W, W')$, where $W=x/\sqrt{\beta}$, $W'=(\alpha x+\beta x')/\sqrt{\beta}$. In such a normalized phase space, particle trajectories form concentric circles of different radius as shown in Fig. 2. When the dispersion kick $(x_{kick}, x_{kick}')$ is normalized, $(W_{kick}, W_{kick}')=(0, 2\sin(\theta/2)\sqrt{\beta})\delta$, and for the sake of this simplicity, the direction of dispersion kick in a sector dipole always stick upward no matter what Twiss parameters are in this dipole.

The motion of a particle in a double bending cell can be characterized as follows: A particle moving along a circle line get a kick from the first dipole, then it moves along a new circle, after an angle of the phase advance between the two kick points a kick from the second dipole moves it into the path of a third circle. If after two such kicks, the particle goes back to its original circle path, the cell will work as if no dispersion exist, and it is achromatic. As shown in the right figure of figure 2, from elementary geometry, we can get the achromatic condition: 1. Phase advance between the middle points of the two dipoles must be $\pi$ or $(2n+1)\pi$. 2. The two dispersion kicks in normalized phase space must equal to each other, that is, $\sin(\theta_1/2)\sqrt{\beta_1} = \sin(\theta_2/2)\sqrt{\beta_2}$. These achromatic conditions can be verified by writing down the matrix of each part of the cell and solve the achromatic equation, which is straightforward but involved.

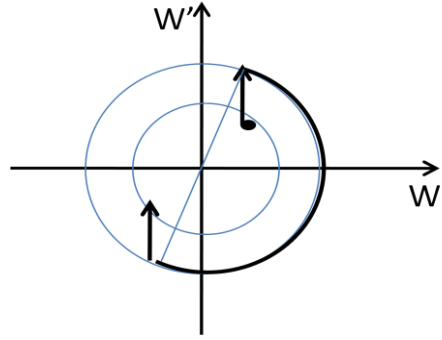

Fig. 2. A schematic figure of the particle in the normalized phase space, it is obvious that two identical kicks with p phase advance move the particle back in its original path, and cancel the dispersion in the cell.

### 2.2 Kick description of the CSR effect

According to R. Hajima's R-Matrix, transverse coordinate change due to CSR effect can also be treated by a matrix as that due to dispersion effect, a sector dipole can be described by a matrix:

$$R_d = \begin{pmatrix} \cos\theta & \rho\sin\theta & \rho(1-\cos\theta) & \rho(1-\cos\theta) & \rho^2(\theta-\sin\theta) \\ -\sin\theta/\rho & \cos\theta & \sin\theta & \sin\theta & \rho(1-\cos\theta) \\ 0 & 0 & 1 & 0 & 0 \\ 0 & 0 & 0 & 1 & \rho\theta \\ 0 & 0 & 0 & 0 & 1 \end{pmatrix}, \quad (2)$$

Where the R(1,5) and R(2,5) terms are the CSR terms. Then with a similar procedure as that in section 2.1, this CSR induced transverse movement can also be described by a kick in the middle of a dipole. For a sector magnet, the CSR kick is calculated from equation $M_{half}$(kick$_{CSR}$, kick$_{CSR}$')=( $\rho^2(\theta-\sin\theta)$, $\rho(1-\cos\theta)$), suc(kick$_{CSR}$,kick$_{CSR}$')=( $\rho^2(\theta\cos(\theta/2)-2\sin(\theta/2))$, $\theta\rho\sin(\theta/2)$). Unforturnately, the CSR kick description is not that simple as the dispertion kick, since the x part of the kick is not zero. But using this kick description is still much easier than the boring 5-by-5 R-matrix manipulation.

### 3. Cancellation of CSR induced emittance in a DBA cell

In this section, we discuss a optics design technique for the CSR induced emittance suppression within an achromatic cell. For the sake of simplicity, we considered a symmetric DBA cell, but this technique can also be used onto other achromatic cells. A symmetric DBA cell is considered, which is composed of two identical sector dipole and the drifts and quadrupoles between them are also symmetric. The Twiss parameters are chosen to satisfy the condition: β1=β2, α1=−α2, where β1,α1,β2,α2 are Twiss parameters at the middle of the two dipoles separately, and the phase advance between the middle points of the first and the second dipole is chosen to be π, which is required by the achromatic condition.

The normalized phase space is still adopted to analyze the particle motion in such a DBA cell. Adding the CSR into consideration, at each kick point the particle experienced two kicks: a dispersion kick and a CSR kick, and the vector sum of these two kicks is the overall effect, as shown in figure 4. After the first kick point, the particle moves along a circular path in the normalized phase space, until it reaches the second kick point. Similar to the procedure in the cancellation of the dispersion, the effect of the CSR can be cancelled if the particle returns to its original path after those two kicks.

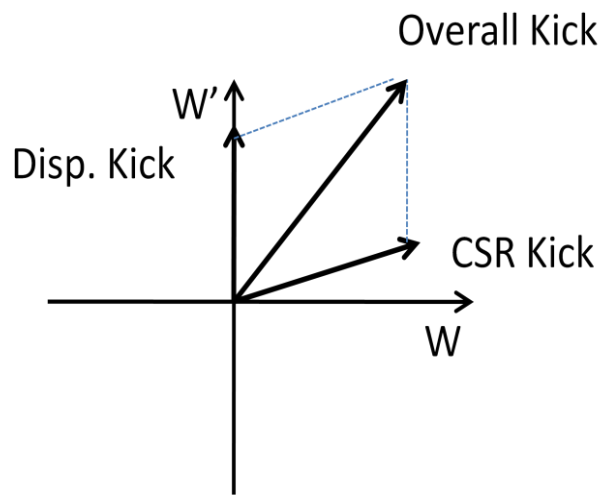

Fig. 4. Dispersion kick and CSR kick at a kick point in the normalized phase space.

For simplicity, we consider a particle which stays at the original point before the first kick. Its trajectory in the normalized phase space will be as follows: It get a kick from the first kick point and its coordinates can be denoted by a vector **R1** in the phase space, then the vector **R1** turn around the original point with a angle π, which is the phase advance between the first and the second kick point, and its vector turns into **R1'**, at the second kick point a second kick vector is added and the position vector turns to be **R1'+R2**. The lattice condition that makes the final position vector **R'+R2**=0 will show a cancellation of the linear CSR in such a lattice cell as shown in Fig. 5..

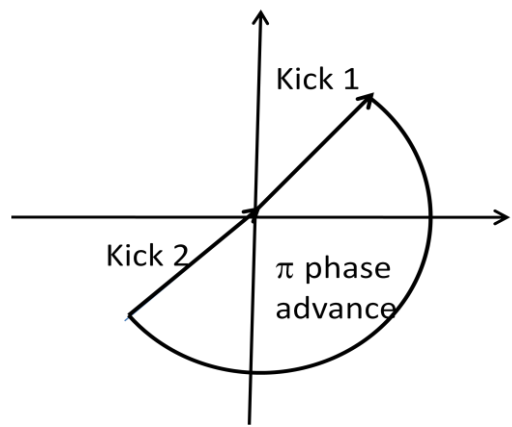

Fig. 5. The particle dynamics in normalized phase space when the CSR and dispersion kicks are considered. Where kick1 and kick2 are all sums of CSR and dispersion kicks, the phase advance is π which is required by the achromatic condition as shown in section 2.1. From this schematic figure, it is obvious that These kick effects cancels when Kick1 = Kick2

After the first kick, the position vector in the normalized phase space becomes:

$$R1 = R_{dispersion}^{norm} + R_{CSR}^{norm}, \qquad (3)$$

Where,

$$R_{dispersion}^{norm} = \left(0 \quad 2\sin\frac{\theta}{2}\sqrt{\beta_1}\right)\delta_1, \qquad (4)$$

$$R_{CSR}^{norm} = ((\rho^2(\theta\cos\frac{\theta}{2} - 2\sin\frac{\theta}{2})/\sqrt{\beta 1}, (\alpha 1(\rho^2(\theta\cos\frac{\theta}{2} - 2\sin\frac{\theta}{2}) + \beta 1(\theta\rho\sin\frac{\theta}{2})/\sqrt{\beta 1}) \qquad (5)$$

Since the phase advance equals π, R1' before the second kick point turns into:

$$R1' = -R1 \qquad (6)$$

And the kick vector at the second kick point is just one that resembles R1, we get:

$$R2 = R_{dispersion}^{norm} + + R_{CSR}^{norm} \qquad (7)$$

and

$$R_{dispersion}^{norm} = (0, 2\sin(\theta/2)\sqrt{\beta 2})\delta_2 \qquad (8)$$

$$R_{CSR}^{norm} = ((\rho^2(\theta\cos\frac{\theta}{2} - 2\sin\frac{\theta}{2})/\sqrt{\beta 2}, (\alpha 2(\rho^2(\theta\cos\frac{\theta}{2} - 2\sin\frac{\theta}{2}) + \beta 2(\theta\rho\sin\frac{\theta}{2})/\sqrt{\beta 2}) \qquad (9)$$

But still some caution is needed, as a result of the CSR effect the energy spread δ2 at kick point 2 is larger than that at kick point 1:

$$\delta 2 = \delta 1 + \kappa\rho\theta \qquad (10)$$

The CSR cancellation condition requires that:

$$R2 + R1' = R2 - R1 = 0 \qquad (11)$$

Substituting equation (2),(6) into the equation above, and considering that β1=β2, α1=−α2, the CSR cancellation condition in a DBA cell can be get as:

$$\beta_1 = \frac{\alpha 1\rho(-2 + \theta Cot\frac{\theta}{2})}{\theta} \qquad (12)$$

for θ<<1, which usually can be satisfied in a transport line, this condition can be simplified as:

$$\beta_1 \sim \alpha 1\rho\theta/6 \qquad (13)$$

## 4. Particle Numerical Simulations to verify the emittance suppression

This method of emittance compensation is confirmed by a simulation using the particle tracking code ELEGANT. The initial condition of the electron bucn is assumed to be: central energy 1GeV, bunch charge Q=500 pC, normalized emittance $\varepsilon_n$=0.2 mm.mrad, bunch length σs=100fs.

For our purpose, symmetric DBA cells with different initial Twiss parameters a constructed and simulated using ELEGANT, bending angle of dipoles are 3 degree. From our emittance cancellation condition (10), we can find the requirement on the Twiss parameters at the entrance of the DBA cell:

$$2\beta_0^2 \rho \cot(\theta/2) - 2\beta_0 \alpha_0 [\rho^2 - \cot^2(\theta/2)] - 2\rho \cot(\theta/2)(1+\alpha_0^2) = 0. \quad (14)$$

Where $b_0$, $a_0$ are Twiss parameters at the DBA entrance, and q, r are bending angle and radius of the dipoles. Simulation results are shown in Fig. 6, where different colors denote different emittance growth. A redder color means a larger emittance growth. The theoretical condition of (12) and that of envelope matching are also shown in the figure. It is shown that the DBA cells with original twiss parameters which satisfies our emittance cancel condition really gives the smallest emittance growth, and our condition is better than that of envelope matching.

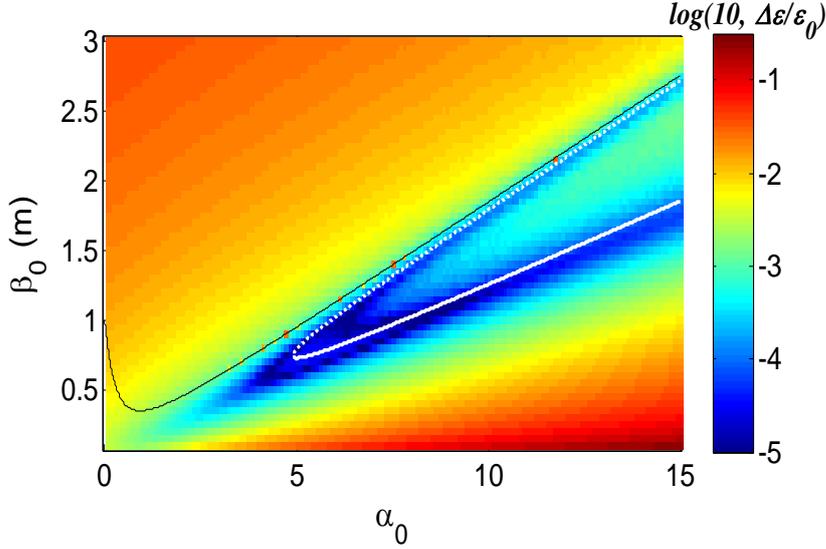

Fig. 6. Simulation results of DBA lattices with different α0, and β0. The white line represents our condition (12), and the black line represents lattice satisfy the envelope matching condition.

## 5. Conclusion

In conclusion, we have derived a optics design technique which can significantly suppress the emittance growth induced by CSR in a DBA cell, and this has been verified by simulation results. We believe that using similar derivations shown above, this technique can be expanded to use in the design of other types of achromatic cell such as asymmetric double bending cells or TBAs, etc. However, in our study above we have assumed a constant bunch length and linear approximation of the CSR induced energy spread, so that in beam transport systems where the bunch length changes significantly, or nonlinear effect of CSR takes an important role, further studies are still needed.